# Better Decision Making in Drug Development Through Adoption of Formal Prior Elicitation

Nigel Dallow[1], Nicky Best[2], Timothy Montague[3].

[1]GlaxoSmithKline R&D, Clinical Statistics, Uxbridge, UK
[2]GlaxoSmithKline R&D, Statistical Innovation Group, Uxbridge, UK
[3]GlaxoSmithKline R&D, Clinical Statistics, Upper Providence, USA

*Abstract*

With the continued increase in the use of Bayesian methods in drug development, there is a need for statisticians to have tools to develop robust and defensible informative prior distributions. Whilst relevant empirical data should, where possible, provide the basis for such priors, it is often the case that limitations in data and/or our understanding may preclude direct construction of a data-based prior. Formal expert elicitation methods are a key technique that can be used to determine priors in these situations. Within GlaxoSmithKline (GSK), we have adopted a structured approach to prior elicitation based on the SHELF elicitation framework, and routinely use this in conjunction with calculation of probability of success (assurance) of the next study(s) to inform internal decision making at key project milestones. The aim of this paper is to share our experiences of embedding the use of prior elicitation within a large pharmaceutical company, highlighting both the benefits and challenges of prior elicitation through a series of case studies. We have found that putting team beliefs into the shape of a quantitative probability distribution provides a firm anchor for all internal decision making, enabling teams to provide investment boards with formally appropriate estimates of the probability of trial success as well as robust plans for interim decision rules where appropriate. As an added benefit, the elicitation process provides transparency about the beliefs and risks of the potential medicine, ultimately enabling better portfolio and company-wide decision making.

## 1. INTRODUCTION

Within the pharmaceutical industry, Bayesian methods are increasingly utilised for statistical analyses or to aid design of future studies (e.g. via the assessment of assurance [1] or decision-criteria operating characteristics). One key challenge when using Bayesian methods is the selection of the prior. In some instances data in a similar setting has been previously generated and can be used directly to construct a prior. However, there are generally differences between trials (e.g. different populations, endpoints, durations etc.), the impact of which can be difficult to quantify directly, and additional data may exist (from other compounds or other indications, pre-clinical data etc) that would be problematic to formulate mathematically into a prior. In other settings (e.g. novel mechanism of action, novel endpoint) there may be little or no relevant data with which to directly formulate a prior. One approach to deal with these situations is to draw on expert knowledge and experience to "translate" the available information into a prior distribution. At GSK we have adopted a formal prior elicitation framework proposed by Oakley and O'Hagan [2] known as SHELF. It has now become routine for teams to use this approach in conjunction with the calculation of assurance [1] at key decision milestones. Here the probability of success (assurance) of the next study is determined by weighting the conditional likelihood of success (e.g. power) for fixed effect sizes by a prior distribution for the true value of the treatment effect of interest.

The aim of this paper is to share our experiences of adopting a formal prior elicitation process at GSK, which we have applied to over 30 studies during 2015-16. We will first provide a brief introduction to the SHELF prior elicitation framework (Section 2) followed by a discussion of the benefits (Section 3) and challenges (Sections 4 and 5) of conducting prior elicitations, and our experiences of how to structure the quantities to be elicited (Section 6). We then present a series of case studies to highlight some specific benefits and challenges (Section 7). Finally, we provide some additional thoughts on our experiences with elicitation as well as comment on how elicitation is being utilized elsewhere (Sections 8 and 9). Additionally, a companion paper [3] in this journal provides similar details of how GSK uses a standardised assessment of assurance for all key decisions on investment for future trials.

## 2. PRIOR ELICITATION FRAMEWORK

Elicitation is an interview process in which experts are asked a series of questions about their beliefs regarding one or more uncertain quantities (e.g. about a mean treatment effect). Based on each expert's responses to these questions, the statistician will then derive a probability distribution for the quantity of interest that reflects what the expert believes about the value of the quantity as well as the uncertainty of that belief. The SHELF process developed by Oakley and O'Hagan [2] uses a structured approach to prior elicitation that is summarised in Table 1.

**Table 1: Main steps in SHELF elicitation process**

| 1. Select experts | These can be both internal and external to the company and should involve only those that have a good understanding of the details that need to be elicited. |
|---|---|
| 2. Train experts | Provide experts with an overview of the elicitation process and the use of subjective probabilities and probability distributions |
| 3. Evidence dossier | Prepare and review an evidence dossier that captures all pertinent information that the experts would rely upon to formulate their opinion. |
| 4. Elicit individual priors | Elicit, in a masked fashion, individual priors from each expert (i.e. experts are unaware of what other experts believe at this point) |
| 5. Discuss individual priors | Share and review results from individual elicitations including each expert's rationale for their beliefs; discuss differences between experts. |
| 6. Agree consensus prior | Where possible, elicit a 'consensus' prior from the experts which is based on what they collectively agree a 'Rational Independent Observer' would determine after having observed the previous conversations. |
| 7. Documentation | Provide a written record of the elicitation session |

For both the elicitation of the individual priors and the consensus, different techniques are defined in SHELF including elicitation of tertiles or quartiles, roulette and probability based elicitation. We have used both quartiles and roulette methods to elicit individual priors, with the latter being our preferred approach as experts find it more intuitive to use. To elicit the consensus prior, we use the probability method, as we have found it more conducive for a group discussion. A brief description of the three methods are presented below; see [2] for more details.

The *quartile* method requires experts to provide the $25^{th}$, $50^{th}$ and $75^{th}$ quartile values of their belief about the true value of the unknown parameter. The *roulette* method requires experts to use "chips" to build a histogram to represent their beliefs about the true value of the unknown parameter. The *probability* method requires experts to provide probabilities that the true value of the unknown parameter is less than each of three distinct values (e.g. $X1 < X2 < X3$).

During the elicitation process it is critical to provide real-time feedback to the experts regarding their individual beliefs/priors as well as the resultant consensus prior. As part of the SHELF procedure, an R Package [4] has been developed that provides real-time visualisation of both individual and consensus elicited curves by selecting the parametric probability distribution (e.g. normal, log-normal, etc) which most closely matches the elicited values. A key feature of the SHELF approach is the role of the *facilitator* who has expertise in the process of elicitation. The facilitator guides the experts, manages the process to ensure that all viewpoints are shared and debated, and at the end delivers the fitted probability distribution(s) representing the experts' beliefs. The facilitator has a critical role in navigating the numerous challenges that can arise during an elicitation. Elicitations carried out at GSK typically have two facilitators, one to lead the overall process and one to record key details of the elicitation session and run the software to fit probability distributions to the values elicited from the experts.

## 3. USES AND BENEFITS OF PRIOR ELICITATION IN DECISION MAKING

In addition to quantifying expert knowledge into a probability distribution that can be utilized in a Bayesian framework, we have found multiple other benefits of adopting a formal prior-elicitation process.

### 3.1. More transparent calculation and communication of Probability of Success

At GSK, at key milestones in development, internal scientific and investment boards review both the scientific and financial plans for teams seeking investment to the next stage of drug development, where information on the probability of success for both the next trial and development plans as a whole is considered. To aid these reviews, teams are expected to discuss with these governance boards details of assurance estimates together with the underlying prior distribution, for all key milestones. Underpinning this is the choice of prior, which is frequently determined via prior elicitation. The elicitation process highlights not only the rationale for believing in the likely effect of the drug, but the gaps in knowledge and/or sources of uncertainty. Ultimately this has enabled more robust portfolio decisions to be made and, where necessary, led to changes in development plans where risk mitigation has been needed as a consequence (e.g. agreement for staged investment or adoption of futility rules in the next trial).

## 3.2. Improvements in Study Designs

Discussion of the evidence dossier and each expert's individual prior distribution may reveal aspects of the study design that could compromise the robustness of the study, and which had not been considered previously. For example, different beliefs about a treatment effect may be driven by opinions on target patient population or other design factors such as use of rescue/concomitant medication (see Case Study 2 for an example). Prior elicitation provides a mechanism by which teams are able to identify these issues and their potential impact on study success, and to subsequently address them in the study protocol.

In cases where the elicited prior represents a reasonable probability that the treatment effect will be negligible or not clinically meaningful, teams may consider study design features (e.g. interim analyses) and/or clinical development plan options (e.g. conducting studies in parallel or sequentially) to help mitigate key uncertainties (see Case Study 1). The elicited prior distribution can be used to assess various study designs (e.g. number of interims and sample size at interims) and decision rules, in order to find the design that provides the optimal statistical operating characteristics (e.g. probability of making a correct 'go' or 'no go' decision).

More generally, elicited priors have been used extensively at GSK in clinical trial simulations to assess operating characteristics of different trial designs, such as to calculate series of assurance estimates to assess the impact of different sample sizes and alternative definitions of the end of study success criterion, or to compare overall probability of success for different testing hierarchies when multiple endpoints are of interest. See [3] for some further examples.

## 3.3. Deeper Understanding

Perhaps most importantly, we have found that the prior elicitation process has facilitated rich and scientifically-driven reviews of evidence, enabling more robust collective understanding and decision making. Although experts have the same data and information in front of them, they often independently formulate different conclusions. Formal, facilitated prior elicitation sessions have enabled the experts to have deeper discussions about the existing knowledge and data, and to probe to understand why some experts may place more weight on certain evidence compared to other experts. This allows the experts to identify any gaps in their knowledge and/or sources of uncertainty, leading to the experts debating the evidence robustly. Previously, study teams at GSK were having some discussion but not to the extent now routinely achieved during a formal prior elicitation session, and they were generally not critiquing evidence as deeply.

## 4. TECHNICAL AND STATISTICAL CHALLENGES WITH PRIOR ELICITATION

There are many technical and statistical challenges which can lead to bias in prior elicitation sessions [*5*]. It is important to be aware and proactively address these when running the elicitation process. The use of the SHELF protocol can help with this. Here we highlight some specific challenges that we have encountered which require careful consideration, and propose some strategies and solutions to minimise or overcome these.

## 4.1. Aspiration vs belief

Biases can arise particularly in situations where data are sparse. For example, in early-phase clinical studies where limited or no data exist, we have found that without careful guidance experts can provide priors representing treatment effects that they *want to observe* rather than what they currently believe the true effect to be. Experts often struggle with the concept of eliciting the 'true' treatment effect and risk bringing sampling uncertainty and aspirational beliefs into the elicited prior. Thorough training of experts is critical to minimise such risks. The facilitator needs to concentrate on the language of the experts and the scientific rationale for their beliefs, and challenge the experts if necessary to ensure the 'what they want to see' aspect does not creep into the process.

## 4.2. Over-optimism

Careful selection of experts is paramount to a successful prior elicitation session. Many of the experts are likely to be either Key Opinion Leaders (KOLs)/treating physicians or members from the project team. In this situation the facilitator needs to ensure that the experts do not either intentionally or unintentionally give over-optimistic views. One aspect of the SHELF process is to document any potential sources of known biases (e.g. conflicts of interest) to create transparency. The facilitator should also ensure that experts justify their opinions during the feedback and discussion following the individual prior elicitations. Benchmarking the expert's opinion with 'portfolio priors' can also be helpful. Simply sharing with the experts the overall success rates for assets at a particular stage of development within either the disease area or the broader industry level (e.g. [6]) can help experts better calibrate risks of novel mechanisms not translating to clinical efficacy.

A related problem is that standard elicitation approaches require experts to provide a *uni-modal* distribution. Experts may have some belief that the true drug effect could be near zero but think it is implausible for it to be negative/unfavourable. As a result, experts may tend to push their distribution towards more positive results to avoid parts of the fitted distribution being negative. Fitting a bounded or truncated distribution can help, but this can still fail to adequately capture the expert's beliefs about the most likely values whilst still having sufficient probability of near-zero (or negative) effects. We have found that a better solution can often be to elicit a bi-modal prior distribution (see section 6.3 and Case Study 3).

## 4.3. Challenges reaching consensus

The SHELF process defines what is called a 'consensus' prior, using a behavioural aggregation approach [2]. Following the elicitation and discussion of each expert's individual prior, experts are then asked to collectively agree a consensus prior representing what a 'Rational Independent Observer' (RIO) would determine after having observed the previous conversations. This requires the experts to implicitly agree on how much weight should be given to each of their individual priors and supporting arguments; it does not (and should not) require each expert to accept the RIO consensus as their own personal prior belief. However, in some instances we have found that either experts cannot put aside their beliefs or there are fundamental and valid differences in opinions. Ultimately, failure to reach a consensus has not proved to be a major issue, provided the elicitation exercise can formulate *why* differences occur. In these situations we have found that eliciting two (or more) separate priors based on the differing opinions and providing a clear supporting rationale can be very helpful to decision makers and governance boards, so that open and transparent review of risks can then be held based on the feedback from the elicitation - see Case Study 2 (section 7.2) for an example.

Alternatively, a consensus prior can be derived by mathematical pooling of the experts' individual priors. We do not recommend this in situations in which there are strong differences of opinion between experts, since a mathematical average can end up representing no-one's beliefs in particular. However, when there is a high degree of overlap of all the experts' individual priors, mathematical pooling (using a simple average of the individual priors) can be an efficient and acceptable method to achieve a consensus prior. In such situations, after first explaining the purpose of the consensus prior to the experts, we may choose to show them the mathematically pooled prior and ask if they all agree that this provides a reasonable representation of what an independent observer might believe.

### 4.4. Risk of experts misunderstanding statistical quantities

To run a successful elicitation process the experts require a good understanding of both probability and other statistical terms (e.g. quartiles). When we originally piloted prior elicitation, we started with the quartile approach for individual elicitations. Here we required the experts to give their median, lower 25% and upper 75% quartiles for the true value of the quantity of interest. It was clear that some experts didn't have a good conceptualisation of these terms, leading at times to 'U' shaped distributions which, when reviewed, did not in any way represent their beliefs. Through careful education using an initial training session we were able to overcome some of these misunderstandings; however we still at times encountered problems with distributions not truly matching the expert's belief. Partly for this reason, our preferred method for eliciting individual expert priors is the roulette approach, which experts find more intuitive and is less prone to misunderstanding.

Another issue is that experts often struggle with the concept of eliciting the 'true' treatment effect and risk bringing in sampling uncertainty into the elicited prior. Clarity in the training of experts to ensure that they understand what is meant by 'true effect' (e.g. result of the *infinitely sized* clinical trial) and other statistical concepts is critical to minimise such risks.

### 4.5. Risk of experts providing symmetrical 'bell-shaped' distributions

Although there is nothing wrong with an expert providing a symmetric 'bell-shaped' distribution for their beliefs, we have found that experts often feel that this is required, having been continually presented with normal distributions during statistical training. When using techniques such as the roulette method we are careful to ensure that experts understand that their distributions do not have to be symmetrical (by highlighting both symmetric and non-symmetric examples during training) and should truly reflect their beliefs. The role of the facilitator is also extremely important here: during the elicitation, he/she should ask questions around each expert's prior – focusing not only where the distribution is centered but also on the tails of the distribution – to ensure it truly matches the expert's belief.

# 5. PRACTICAL AND LOGISTICAL CHALLENGES WITH PRIOR ELICITATION

## 5.1. Choice of experts

One key aspect of a prior elicitation is the selection of experts. Individuals with the relevant experience and expertise are often members of the project team or KOLs who may have a vested interest in the running of the trial (e.g. unmet need for patients). Although there is nothing wrong with such individuals being experts for the elicitation process, there is an expectation of unconscious bias based on their experiences and backgrounds with the potential to lead to over-optimism. In sessions we have run, we have tried to be comprehensive in the selection of experts, aiming to get at least some experts (internal or external to the company) who are independent of the project if possible. We also typically include a statistician who has in-depth knowledge of the project or related development programmes; he/she brings both a detailed understanding of the relevant data and expertise in expressing judgements and uncertainty in the form of probabilistic statements. There is no ideal solution to the balance between expertise and impartiality when selecting the experts, and having a skilled facilitator is critical to ensure experts' in-depth knowledge is brought to the prior elicitation process effectively in an unbiased fashion.

## 5.2. Complexities of running elicitation sessions when experts are located remotely.

It is common that experts work internationally across multiple continents and this poses logistical challenges for running an elicitation. Given the psychological aspect of prior elicitations, face-to-face interactions are preferable. Therefore, we have tended to utilise video-conferencing to try to engage virtual face-to-face elicitations. Additionally, we have had an experienced facilitator at each location to assist the lead facilitator in managing and 'reading' the room.

## 5.3. Evolving project planning

In reality, developing study plans can be iterative in nature, with changes to study design occurring until late in the protocol development. When using prior elicitation to determine estimates of assurance, all key aspects of the design need to be well defined. Even during the elicitation session itself, key facets of the design that impact on the expected treatment effect can arise which may necessitate adapting the details of what is to be elicited.

One particular observation is that experts may disagree with the populations proposed and believe that with a different population the probability of success could increase. It is critical that all experts have a precise understanding of the study design to which the elicitation relates (including targeted population) and if changes are needed they are made before the experts give their priors. For example, during the review of the evidence package for an elicitation to support a proposed paediatric study, one of the experts (external to the project/study team) noted that, based on the targeted population, the likelihood of the treatment effect being positive was unlikely and a more uncontrolled population would significantly increase likelihood of success. Following a discussion, the experts agreed that the drug needed to be studied in a more uncontrolled population, and so the experts were asked to provide their priors for the true treatment effect in this new population.

## 5.4. Software

It is important to be able to provide real-time feedback on both the individual and consensus priors. The R package [4] provided as part of the SHELF protocol has been a very useful tool, which we have enhanced by building a SHINY interface [7]. The interface provides faster and improved graphical representation of both individual and consensus priors, and has been tailored for the process of recording the information elicited from experts. Additionally, the tool has extra outputs that we have found to be useful, including presenting elicited priors on different scales (e.g. presenting the implied prior for active arm response if the elicited quantities are the control arm response and the treatment difference; presenting elicited priors for active and control rates on a hazard ratio scale or vice-versa).

# 6. STRUCTURING THE ELICITATION PROBLEM

## 6.1. Defining the quantities to elicit

Before conducting an elicitation, the facilitator needs a clear understanding of the statistical model or decision problem of interest and the uncertain quantity(ies) for which an elicited probability distribution is required. Often the parameter of interest is a measure of the treatment effect relative to the control. This could be the absolute difference between the treatment and control, a relative treatment difference or the hazard ratio (i.e. time to event outcome).

Having established the statistical quantities of interest, it is then necessary to determine what exactly the experts will be asked to elicit. The elicitation variables should be defined in such a way that the expert is able to apply his/her knowledge as directly and fully as possible without necessitating 'mental gymnastics'. This can vary between experts, and can depend on the nature of the available evidence that the experts are drawing on. For example, for an elicitation carried out to inform a cardiovascular outcomes trial, the main source of evidence comprised several previous trials reporting the hazard ratio for the same endpoint for competitor molecules. Experts therefore felt comfortable directly eliciting their beliefs about the hazard ratio. On the other hand, in an elicitation for a rare disease with a novel endpoint, no previous comparative studies were available and the only relevant evidence was on disease progression rates from a natural history study, plus some limited PK-PD data on the molecule of interest. We therefore chose to elicit experts' beliefs about: (1) the proportion of patients who would progress by 18 months on placebo; (2) the (relative) difference in this proportion between active and placebo. Under the assumption of exponential event times, these two quantities together determine an implicit prior for the hazard ratio that can be used for assurance calculations [8].

## 6.2. Eliciting correlated quantities

In situations where we are eliciting expert beliefs about both control and active responses, it is often the case that what an expert believes about the response on treatment will depend on what he/she believes about the control response. In such situations, it is important to recognise this potential dependence when structuring the elicitation problem. One way to address this is to ask the experts to assume a particular value for the control response (typically, the mean or mode of the elicited control prior), and then elicit their beliefs about the active response conditional on the fixed control value. This conditional distribution can then be converted into a prior for either the relative or absolute treatment difference between active and control, and then combined with the prior for the control response to derive an unconditional (but correlated) prior for the active response. Alternatively, we often elicit the experts' beliefs about the absolute or relative treatment difference directly. The choice of absolute or relative difference will depend on which measure of treatment effect the experts think is most independent of the control response, and this will be discussed and agreed with the experts at the start of the elicitation.

Another scenario involving correlated quantities is elicitation of parameters for dose-response modelling. For example, we may require prior distributions for the Emax, E0 and ED50 parameters of the three-parameter Emax model [9], which are known to be highly correlated. Additionally, experts may be unfamiliar and uncomfortable providing beliefs about these parameters. Rather than directly elicit expert beliefs about each of these parameters and their dependencies, we follow an approach broadly similar to that of Huson and Kinnersley [10], which is to elicit expert beliefs about the true response for each of a pre-specified set of doses of interest. We then map these beliefs onto the dose-response model parameters by treating the elicited prior for each dose as if it was the sampling distribution for the response at that dose and fitting an appropriate Bayesian dose-response model assuming functional uniform priors [11]. The resulting joint 'posterior' distribution for the model parameters represents the implied expert belief distribution for those parameters. Case study 4 in Section 7.4 provides an example.

For all of the above situations, it is useful to be able to derive and show distributions of the final parameter of interest (e.g. hazard ratio, absolute difference, mean dose-response curve etc) to the experts to ensure it represents their beliefs. We have adapted our software for this purpose (section 5.4).

### 6.3. Eliciting bi-modal prior distributions

In many situations there is a reasonable probability that the drug being tested will demonstrate no efficacy in the planned endpoint of interest. Furthermore, it may be scientifically implausible that the drug being tested will have a true negative effect. As noted in section 4.2, in this situation eliciting a uni-modal distribution may not accurately capture an expert's belief about the true effect since there may be insufficient probability around a near-zero effect.

To overcome this issue we have, in a number of elicitations, elicited bi-modal ('spike and smear') distributions. We do this by 1) eliciting the expert's probability – $w$, say - that the drug has a true positive/favourable effect, and 2) eliciting the distribution of this effect size under the assumption that the drug *does* have a favourable effect. We then form a mixture distribution to represent the overall prior for the treatment effect. For example, if the prior conditional on the drug having a favourable effect has been elicited on the scale of the treatment difference, we then weight this distribution by a factor (1-$w$) and add a 'spike' with weight $w$ at zero (absolute difference) or one (relative difference) to represent the probability of no effect. Figure 1a gives a hypothetical example to illustrate this approach. If a prior for the control response has also been elicited, this can be combined with the bimodal prior for the treatment difference to provide a mixture prior for the active response which has weight $w$ on the control response distribution and weight (1-$w$) on the conditional distribution for the active response (Figure 1b).

**Figure 1: (a) Illustration of bi-modal 'spike and smear' distribution for a treatment difference. In this case, experts have given 40% probability that there is no true benefit of the drug (represented by spike at zero) and therefore 60% weight to the elicited prior conditional on a positive treatment effect. Note that due to the difficulty in plotting a mixture of a discrete and continuous distribution (the height of the spike is ill-defined), we follow Walley et al [12] and scale the height of the spike to equal its probability mass (0.4) and scale the continuous conditional distribution so that its maximum equals its total probability (0.6). (b) Illustration of mixture prior for response on active treatment. Blue distribution is the elicited prior for the placebo response; red distribution the elicited prior for the active response conditional on drug having a true benefit; black distribution is the overall distribution for the active response and is a mixture of the blue and red distributions with weights 40% (the experts' consensus probability for there being no true benefit) and 60% respectively.**

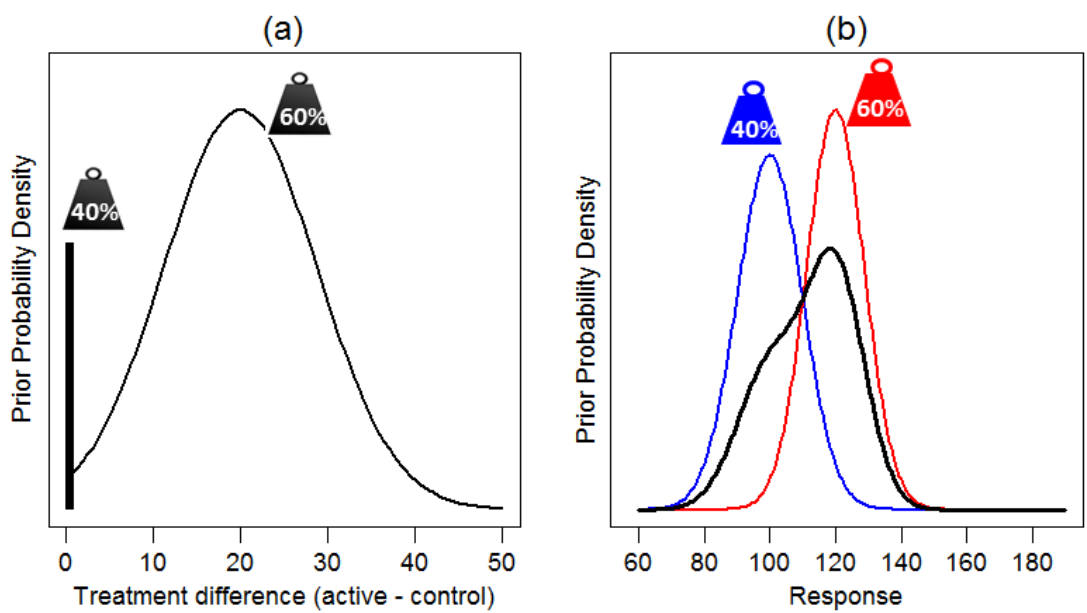

A useful source of information when eliciting bi-modal priors are statistics on the success rates in drug development [6]. Through the use of cross-industry benchmarking of success at each stage of drug development we are able to provide experts with background success rates for many diseases under consideration for compounds at different stages of drug development. These statistics can help calibrate experts' opinions on the probability of a compound failing to demonstrate efficacy. Walley et al [12] used such a 'portfolio prior' in a case study in which they elicited a prior for the treatment effect in a Proof of Concept study. Their approach was to construct a bi-modal – or "spike and smear" – prior by eliciting a distribution based on expert beliefs about the treatment effect if the compound was 'active', and then adding a spike of probability at zero with weight 60%, which they argue is comparable with industry attrition rates at this stage of development. In contrast, we have found it more useful to use these portfolio success rates as additional background evidence, to be taken together with all information on the specific compound of interest that the experts synthesize to formulate their prior. This forces experts to robustly articulate and justify if and why they believe the probability of a true positive effect is higher for a specific compound than the overall portfolio success rate, which in turn is valuable information for decision makers and investment boards when balancing risks across the portfolio.

# 7. CASE STUDIES

During 2015-2016 we have conducted over 30 formal prior elicitation sessions using the SHELF approach. Elicitations have been run across a variety of therapeutic areas and all phases of drug development, with a reasonable balance in numbers across Phases II-IV, plus a few at Phase I stage. From these we have selected four case studies that emphasise different aspects of both the benefits and challenges with prior elicitation sessions.

## 7.1. Case Study 1: Rhinitis study

This elicitation focussed on a fixed-dose combination (FDC) of two different drugs with different mechanisms of action. Positive data were reported in a phase II Proof of Concept (PoC) study in an allergen challenge model in which rhinitis patients were exposed to controlled amounts of allergen and the effect of the FDC was assessed using symptom scores. The next planned study was to assess the FDC in a phase III study assessing rhinitic patients in a real-world environmental setting. In addition to the PoC study, data were also available from other similar in-house molecules assessed both in the challenge model as well as an environmental setting, plus summary results from a similar FDC in a series of phase III trials. This set of data allowed the degree of association between phase III response and PoC response to be assessed.

The team developing the phase III study were required to provide an estimate of assurance to seek agreement with internal governance boards to commit to phase III and then to optimise the phase III trial based on levels of assurance. It was decided that although there was a wealth of available data, there were still uncertainties around what the effect of the FDC would be in the setting of Phase III (i.e. environmental setting), so a prior elicitation session was conducted. A group of 6 experts were convened for the elicitation, and their individual and consensus priors for the treatment difference between the FDC and the corticosteroid component are presented in Figure 2.

In parallel with the elicitation session, a model-based prior was derived based on assuming a linear relationship between the PoC and phase III treatment differences. Intercept and slope parameters describing this relationship were included in a Bayesian hierarchical model which updated vague priors with all available PoC and phase III study results from other compounds. The predicted phase III treatment difference based on the PoC data for the investigational product was then used as the model-based prior. This model-based prior is also shown in Figure 2 (red curve). It is noticeable that this has considerably more uncertainty than the experts' elicited priors. The elicited priors have negligible probability that the treatment effect is less than zero (a scientifically implausible result when comparing combination to monotherapy), whilst the model based approach suggests this is plausible. Conversely, the model-based prior also has considerable probability of a treatment difference above 1.0, in contrast with the elicited priors. A benefit of the elicited priors is the fact that the experts are able to bring in the broader knowledge of the FDC mechanisms and published data on other molecules leading to belief that the effect would be very unlikely to be negative, while clinical expertise of the disease informed a maximal plausible efficacy threshold.

The elicited consensus prior was subsequently used to determine estimates for assurance (approximately 55%), which in turn informed the sample size of the Phase III studies as well as the overall development strategy. As a result of quantifying the risk of failure the development team favoured a staggered approach to the two phase III studies to mitigate the cost and risk of a parallel phase III approach. The model-based assurance (approximately 80%) was over-optimistic, and would have led to a less appropriate development plan. See [3] for further discussion of the assurance calculations for this Case Study.

**Figure 2: Results from Rhinitis elicitation session**

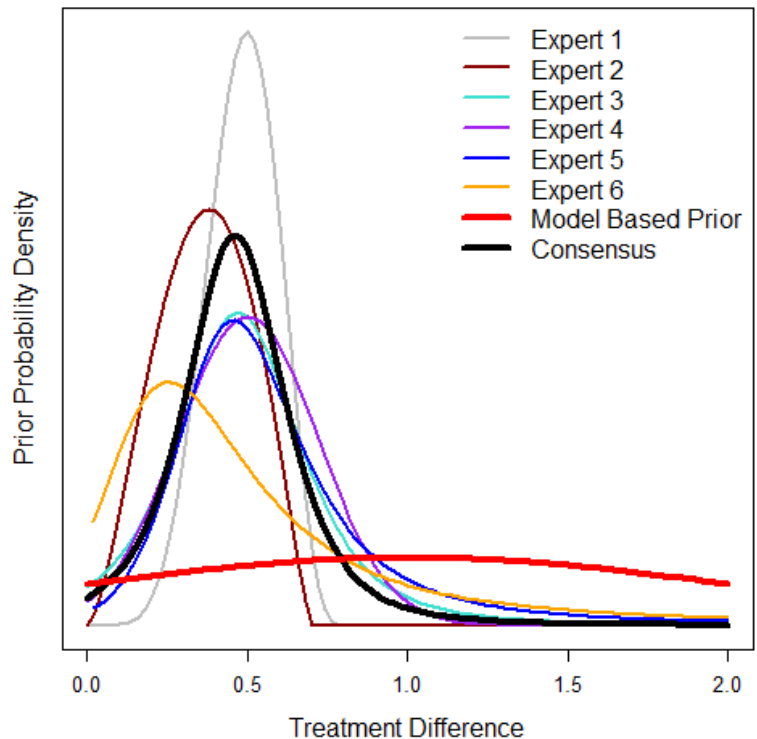

## 7.2. Case Study 2: Rare Disease

A phase III trial was planned in a rare disease, following positive results seen in Phase III for other indications and compassionate use data in the disease area. To determine good estimates of the probability of success (assurance) for Phase III a formal prior elicitation was conducted. Given the rare diseases nature it was decided to utilise external, as well as internal, experts.

The elicitation focused on the primary endpoint of the proportion of patients who experience an exacerbation during the treatment period. The elicitation was structured to first elicit the placebo exacerbation responder rate and then to elicit the relative treatment difference (i.e. relative risk) conditional on the placebo response. The results of the individual priors elicited for placebo are presented in Figure 3, which shows clear disagreement between some of the experts.

**Figure 3: Results from Rare Disease elicitation session**

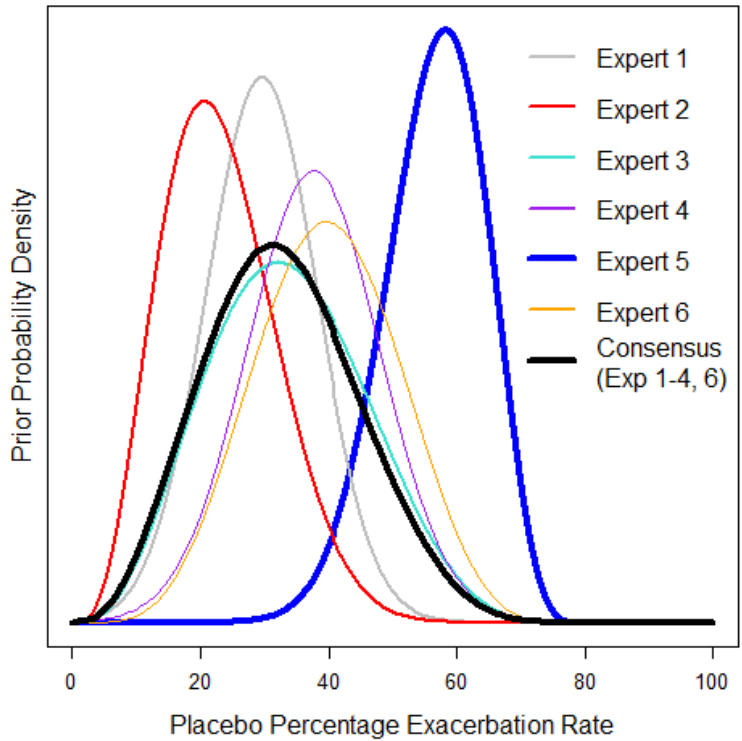

Following review of each expert's elicited prior and discussion around the rationale for their chosen distribution, it was clear that Expert 2 and Expert 5 had fundamentally different beliefs about their placebo response rates. These two experts were both external physicians treating this patient population. During the review of the individual priors, Expert 2 stated that he/she believed that investigators would be keen to get their patients in the trial due to the high un-met need for this disease. Importantly, key inclusion criteria for the study protocol defined that patients must be on a stable dose of background therapy for 4 weeks prior to randomisation and must have experienced two or more exacerbations in the 12 months prior to screening. Expert 2's belief was that investigators who are keen to get patients into the trial, could temporarily reduce levels of background therapy in the months before the trial to induce the required number of exacerbations to meet protocol inclusion, restoring background therapy to the original level for the 4 weeks prior to randomisation and for the duration of the study. Thus, the trial could recruit stable patients unlikely to exacerbate during the study period. In contrast, Expert 5's prior distribution is based on the assumption that investigators, through training, would understand the study endpoint and the target patient population, and hence the population enrolled would be closer to that intended by the protocol inclusion criteria.

Other experts believed that there was some risk of modifying background therapy to ensure patients entered trial, but based on the proposed types of clinical sites planned, they thought the impact on placebo exacerbation rate would be less extreme than did expert 2.

Following lengthy discussion between the experts, the potential of how sites would select patients was felt to be a clear differentiator to placebo response and so it was agreed to have two separate priors for the placebo response. The first prior was the consensus prior based on all individual beliefs *excluding* expert 5's belief. The second was based on expert 5's beliefs.

For elicitation of the relative risk, two separate priors were elicited: one conditional on the consensus prior from all experts excluding expert 5 and one conditional on experts 5's prior.

The rationale for this was to capture priors conditional on how sites would enrol patients and potential risk of background therapy modification. By doing so the experts successfully reached consensus for both of these two priors (not shown here). Based on these priors the team were able to calculate assurance for the phase III trial for each of the two scenarios.

At face value this case study could be seen as a failure to identify a consensus placebo prior. However, ultimately the benefit of the prior elicitation session was that the team were able to have a balanced and transparent discussion at governance boards where quantitative estimates of risks were provided for each scenario, allowing for a better understanding of the risks depending upon how investigators were likely to recruit patients into the trial. Furthermore, the team was also able to adapt the inclusion/exclusion criteria language to minimise the risks identified by Expert 2.

### 7.3. Case Study 3: Secondary Indication

A study was planned in a new indication for a drug already in development. While there was strong biologic rationale for believing the medicine would be effective, there was no prior clinical data in patients for this new indication. As such, an elicitation was planned and a panel of both internal and external experts was convened. Several of the experts had experience treating the target population which comprised patients who failed $1^{st}$ line treatments (often referred to as $2^{nd}$ line treatment). The elicitation was structured to elicit the response of comparator arm (standard of care, SOC), and then the response of the active arm.

The individual priors for the SOC response are presented in Figure 4a (coloured curves), in which it is clear that there are three factions of beliefs about the SOC response. During the discussion of each expert's rationale for their prior it became apparent that experts were considering different patient populations. The two experts (1 and 2) who strongly believed patients <u>would</u> respond to SOC were considering treatment naïve or 1st line patients, based on their personal experience of treating patients in primary care. Conversely, the two experts (3 and 5) who strongly believed that patients <u>would not</u> respond to SOC were considering $2^{nd}$ line patients (who would have failed $1^{st}$ line therapy, SOC) based on their personal experience of treating patients in tertiary referral centres. Realizing the basis for the difference of opinion, the facilitator was able to bring the panel to consensus (black curve in Figure 4a) by clarifying that the target patient population represented those at the more severe end of the disease spectrum who were more likely have an incomplete or no response to SOC.

**Figure 4: Results from Secondary Indication elicitation session. (a) Individual and consensus priors for SoC response rate; (b) Individual and consensus priors for**

active response rate *conditional* on drug working; (c) overall bimodal consensus prior for active response rate.

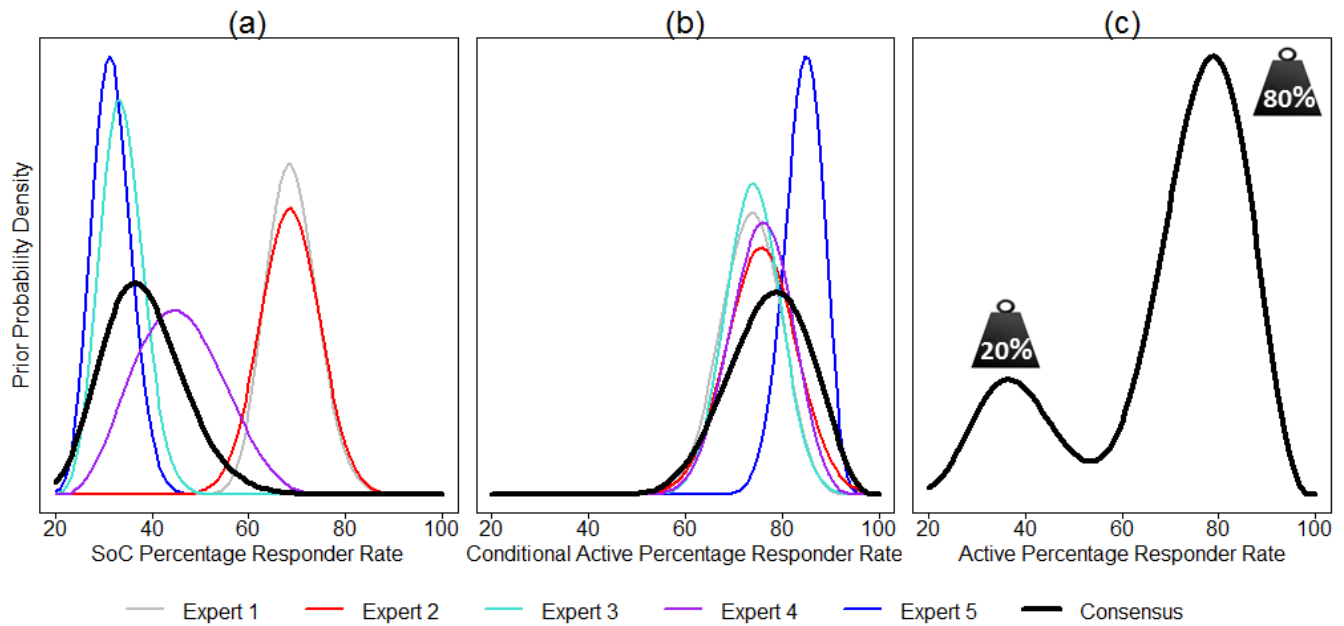

Once consensus had been achieved for the true response on SOC, the experts were asked to provide their beliefs about the true response on the active treatment. Because of lack of data, the experts were first asked to provide their belief regarding the probability that the medicine will provide any level of clinical benefit/efficacy. Then, experts were asked to provide their beliefs regarding the true response of the medicine conditional on the medicine having some benefit/efficacy. Following discussion between the facilitator and experts at the start of the session, it was agreed that the latter would be elicited on the scale of the response rate itself, rather than as a relative or absolute difference in response rates from SOC, because the main evidence related to open-label studies and case reports of response rates in a competitor molecule. Following elicitation of individual priors, the panel was able to agree on both the probability of the medicine providing any benefit (80%), and a consensus prior for the true response conditional on the medicine providing benefit (Figure 4b). In this case, the resulting bi-modal distribution for the response rate on active treatment (Figure 4c) was constructed as a mixture of this conditional distribution (with weight 80%) and the elicited prior for response on SOC (with weight 20%), and was felt by the panel to provide a fair representation of their collective beliefs.

This case study highlights the challenge of ensuring that the quantity to be elicited is clearly defined and that all experts understand that definition. However, through the elicitation process and discussion of rationales for individual beliefs, the facilitator was able to identify the misunderstandings in the definition of the patient population and help the team come to consensus. It should be noted that it is not always possible to reconcile differences in experts beliefs nor is it necessary to do so, as highlighted in the previous case study. The case study also highlights the utility of a bi-modal distribution to allow priors to account for some probability that the drug in question may not be effective.

## 7.4. Case Study 4: Dose Response

The previous examples involved one key treatment comparison of interest. However, in other cases, such as dose finding studies, there may be several treatment effects/parameters of interest. For this case study, the planned study included 5 doses with the main aim to estimate the dose response using a suitable monotonic dose-response model such as a 3-parameter Emax [9] model. In this instance we could have tried to elicit directly the parameters of the Emax model. However most experts struggle to understand and translate these parameters to match their expert judgment of the dose response profile. Also, there was some uncertainty about the most appropriate functional form to select for the dose-response model. Instead, as discussed in section 6.2, the responses at selected doses was elicited and then translated in to a prior distribution for dose response curves of various functional forms. While ideally we would want to elicit at many dose levels, for pragmatic reasons 3 points of the dose response profile (placebo (zero dose), 25 and 50 mg doses) were elicited which allowed us to adequately determine an elicited distribution for a range of 1-, 2- and 3-parameter dose-response profiles. The top dose (50mg) was chosen as the experts believed this would be a dose that gave near maximum effect, and was the top dose planned for the study. The 25mg dose was chosen as this was the predicted equivalent dose to that had been studied in a previous IIa trial (note: there was a change to formulation between Phase IIa trial and planned phase IIb trial). Based on the response distributions elicited at these three dose levels, the experts were then presented with fitted dose-response profiles and associated uncertainty bands for a range of standard models, including linear, log-linear, quadratic, exponential and 3-parameter Emax. Experts were then asked to collectively agree which of the fitted curves most appropriately reflected their beliefs about the likely dose-response profile for the phase IIb setting. The consensus priors for each of the 3 doses and the final elicited dose response profile are given in Figure 5.

**Figure 5: Results from the Dose Response elicitation session. (a) Consensus priors for true response at each of 3 doses. (b) Consensus dose response curve derived from elicited priors**

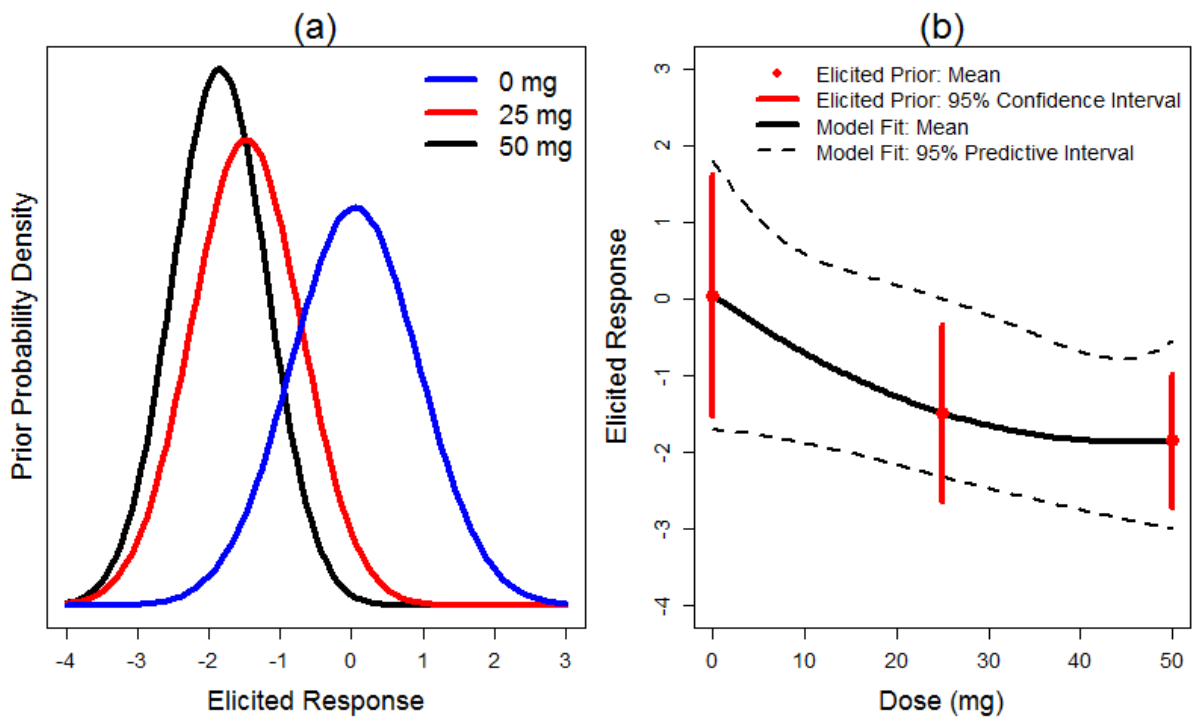

# 8. DISCUSSION

Prior elicitation has become widely used within GSK to derive priors to enable better quantification of success (assurance) and for use more generally in drug development (e.g. in formal Bayesian analyses either at end of study or to support interim results or in simulations to assess different design options). Since instigating an initiative to develop knowledge and expertise in this area in 2014, we have to date conducted over 30 elicitations across many therapeutic areas. Through this journey of learning by implementation we have identified many benefits as well as challenges with prior elicitation. A consistent finding is that prior elicitation allows experts to probe deeply their understanding and beliefs, and to thoroughly discuss why different experts put different weight on different pieces of evidence. By sharing each individual's elicited prior, the discussions become more focussed and quantitative in nature, leading to more robust understanding of risks and, in some instances, insight is gained as to how these risks could be mitigated (e.g. Case Study 2). We have found that use of a formal prior elicitation session stimulates richer, deeper discussions, leading to a tangible benefit over and above the final elicited prior. Teams generally gain a much better appreciation of all available evidence and the nuances associated with this information.

It is important to be able to present complex findings from a prior elicitation session to non-statisticians, including internal governance boards, in a way that is clearly understood. As a result, within GSK we have created a standard template that presents the results of a prior elicitation exercise and associated assurance calculations [3].

One downside of conducting prior elicitation sessions is the time investment required by experts and facilitators. Typically, half a day is needed to perform an elicitation exercise. This could likely involve experts at multiple sites at geographically diverse locations. It is critical that all experts are trained before an elicitation session, which takes approximately 1 hour to complete. To make the process more efficient we are currently developing a e-learning module that allows experts to be trained separately to the prior elicitation session at a time convenient to them.

One key concern commonly raised with prior elicitation is the risk of over-optimism. We have found that, in some instances, the elicitation of bi-modal distributions can help to some extent to mitigate this. Importantly, careful selection and training of experts, structuring and facilitation of the elicitation session and documentation of the session can help to further reduce these risks. Comparing final data to the prior can be envisaged as one way of attempting to review the credibility of the prior, but discordance doesn't necessarily mean that the experts were 'wrong'. Ultimately we will need to compare across many completed studies and assess the overall concordance between priors and final data to start to build a picture of the ability of experts to effectively formulate a prior.

External to GSK, there are several other examples of the use of prior elicitation to inform various stages of drug development. For example, Sabin et al [13] highlight how a quantitative process has been implemented for drugs entering phase II to help support evidence-based decision-making for new drug candidates .The process described shares many of the features of the GSK process described here and in our companion paper [3], including systematic review of the literature and available evidence, use of expert opinion to inform prior distributions on treatment effects, and calculation of probability of success using assurance. However, a key difference is in the way prior distributions are constructed and the role of expert opinion in this process. Sabin et al's approach involves eliciting three separate prior distributions representing sceptical, optimistic and uninformative opinions, and using these to compute three separate assurance values for future phase III studies. This can be helpful to quantify robustness of the decision, but ultimately, one decision, and not three decisions must be made, and so the decision maker must choose which prior they want to believe. The approach we have implemented at GSK enables us to draw on the experts to collectively agree the consensus position to report to decision makers. This is not always possible, as illustrated in Case Study 2, in which case multiple priors and associated assurances are presented. Critically, however, each different prior and assurance value is accompanied by a clear rationale to support the belief, and decision makers can understand the extent to which the experts are truly sceptical or optimistic.

Sabin et al's approach also differs from ours in that they elicit priors for the phase II endpoint, which are then updated with the phase II data once available, and combined where necessary with a model-based prediction of the relationship between the phase II and phase III endpoint in order to provide a new set of priors for the phase III treatment effect. This has the advantage of providing an objective way of incorporating directly relevant data into the prior. However, our experience is that the phase III setting often differs from phase II (e.g. different endpoints, target populations etc.), and there is often insufficient historical data to enable a reliable model-based prediction of the relationship between phase II and III (e.g. Case Study 1). A key motivation for developing the GSK prior elicitation process was to provide a systematic approach to bridging the translational gap between completed studies and planned studies when there is a lack of reliable evidence or scientific consensus, or legitimate models are in conflict.

Walley et al [12] also present a case study to illustrate the benefits of a Bayesian approach to decision making for early phase studies. Their case study relates to a proof of concept trial, but shares many features of the GSK approach which spans all phases of clinical development. In particular, Walley et al also used prior elicitation based on the SHELF protocol [2] to construct informative priors for the control response and the treatment effect. Similar to our experience, their rationale for using prior elicitation rather than a purely data-based prior was that the available historical data were either too limited to provide reliable estimates or related to a different patient population. In addition, they adopt a bi-modal prior for the treatment effect, similar to our Case Study 3, although they base their prior probability of clinically relevant treatment effect on the industry-wide success rate for this stage of clinical development, rather than eliciting this probability from the experts.

One difference between Walley et al's example and our current experience at GSK is that the success criteria in Walley et al's case study were also based on a Bayesian analysis of the trial, whereas the majority of case-studies at GSK have used a frequentist analysis and success criteria. This partly reflects the stage of development, with many of our examples relating to probability of success of phase III designs, where frequentist analyses dominate. As yet, there is limited experience of using elicited priors for analysis of clinical trials. Walley et al distinguish between design priors and analysis priors, and argue that whilst informative priors based on historical data and/or expert opinion can and should be used to inform decision about study design, a more cautious, exploratory approach to using informative priors for analysis is recommended. An exception is the work by Hampson et al. [14], who use expert elicitation to maximise what could be learnt from the analysis of a phase III trial for a rare paediatric disease. In such situations, definitive sample sizes for a clinical trial are usually infeasible, and Hampson et al argue that eliciting expert opinion about likely treatment effects can provide valuable supplemental information that can be updated by results from a prospective clinical trial via a fully Bayesian analysis.

# 9. CONCLUSIONS

High quality decision making is critical to the success of any pharmaceutical company. Whilst relevant empirical data should, where possible, provide the evidential basis for such decisions, decision making in the pharmaceutical industry is often characterized by multiple uncertainties. Prior elicitation is a key tool that allows for capturing current knowledge plus appropriate uncertainty to allow for better quantitative decision making, and enables the wealth of knowledge, experience and insight of scientists and the medical/scientific community to inform such decisions in a rigorous and repeatable way. As an added benefit, the elicitation process provides transparency about the beliefs and risks of the potential medicine, not only by producing a quantitative belief distribution for the treatment effect(s) of interest, but through the process itself, in which all relevant data is summarized, reviewed and implicitly weighted and expert opinions are discussed, debated and documented.

At GSK, putting team beliefs into the shape of a probability distribution provides a firm anchor for all internal decision making: opinions are challengeable and teams are able to provide investment boards with formally appropriate estimates of the probability of trial success as well as robust plans for interim decision rules where appropriate, enabling better portfolio and company-wide decision making.

**Acknowledgements**
We would like to thank Tony O'Hagan for his support and guidance in helping GSK develop expertise in the area of Prior Elicitation. We would also like to thank Adam Crisp and John Davies for their review and input into early versions of this paper and Sam Miller for his input into the Rhinitis case studies.

# REFERENCES

1. O'Hagan A, Stevens J, Campbell M. Assurance in clinical trial design. Pharmaceutical Statistics, 2005; 4: 187–20.1 DOI: 10.1002/pst.175

2. Oakley, J. E. and O'Hagan, A. SHELF: The Sheffield Elicitation Framework (version 2.0), School of Mathematics and Statistics, University of Sheffield, 2010 http://tonyohagan.co.uk/shelf.

3. Crisp A, Miller S, Thompson D and Best N. Practical experiences of adopting assurance as a quantitative framework to support decision making in drug development. Submitted to Pharmaceutical Statistics

4. https://cran.r-project.org/web/packages/SHELF/index.html

5. O'Hagan A, Buck CE, Daneshkhah A, J. Eiser R, Garthwaite PH, Jenkinson, DJ Oakley JE, Rakow T. Uncertain Judgements: Eliciting Experts' Probabilities. Wiley, Statistics in Practice. ISBN: 978-0-470-02999-2

6. CMR International. CMR International Global R&D Performance Metrics Programme, CMR International. *CMRInternational web site* [online], < http://cmr.thomsonreuters.com/ > (2016).

7. https://cran.r-project.org/web/packages/shiny/index.html

8. Ren S, Oakley J. Assurance calculations for planning clinical trials with time-to-event outcomes. Statistics in Medicine (2014). 33(1), 31-45. DOI: 10.1002/sim.5916

9. Kirby S, Brain P, Jones B. Fitting Emax models to clinical trial dose–response data. Pharmaceutical StatisticsVolume 10, Issue 2, DOI: 10.1002/pst.432

10. Huson LW1, Kinnersley N. Bayesian fitting of a logistic dose-response curve with numerically derived priors. Pharm Stat. 2009 Oct-Dec;8(4):279-86. doi: 10.1002/pst.348.

11. Bornkamp B. Practical considerations for using functional uniform prior distributions for dose-response estimation in clinical trials. Epub 2014 ;56(6):947-62. doi: 10.1002/bimj.201300138.

12. Walley R, Smith C, Gale J, Woodward P. Advantages of a wholly Bayesian approach to assessing efficacy in early drug development: a case study. Pharmaceutical Statistics. 2015, 14 205–215. DOI: 10.1002/pst.1675

13. Sabin T, Matcham J, Bray S, Copas A, Parmar M. A Quantitative Process for Enhancing End of Phase 2 Decisions. American Statistical Association (2014) Vol. 6, No. 1. DOI: 10.1080/19466315.2013.852617

14. Hampson LV, Whitehead J, Eleftheriou D, Tudur-Smith C, Jones R, Jayne D, et al. (2015). Elicitation of Expert Prior Opinion: Application to the MYPAN Trial in Childhood Polyarteritis Nodosa. PLoS ONE 10(3): e0120981. doi:10.1371/journal.pone.0120981

15. Kinnersley N, Day S. Structured approach to the elicitation of expert beliefs for a Bayesian-designedclinical trial: a case study (2013). Pharmaceutical Statistics. wileyonlinelibrary.com) DOI: 10.1002/pst.1552